\documentclass[10pt]{article}

\usepackage{amsmath}
\usepackage{braket}
\usepackage{siunitx}
\DeclareMathOperator{\trace}{Tr}
\DeclareMathOperator{\real}{Re}
\DeclareMathOperator{\imag}{Im}
\newcommand{\mat}[1]{\underline{#1}}
\newcommand{\orcid}[1]{~\href{http://orcid.org/#1}{\protect\includegraphics[height=11pt]{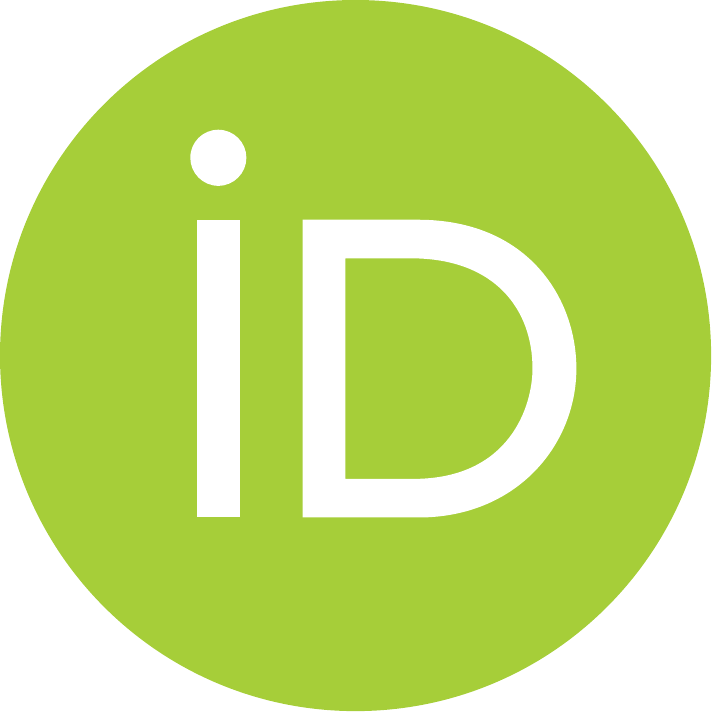}}}

\usepackage[OE]{express}

\begin{document}

\title{Numerical Method for the Maxwell-Liouville-von Neumann Equations using Efficient Matrix Exponential Computations}

\author{Michael Riesch\authormark{*}\orcid{0000-0002-4030-2818} and Christian Jirauschek\orcid{0000-0003-0785-5530}}

\address{Department of Electrical and Computer Engineering, Technical University of Munich, Arcisstr. 21, 80333 Munich, Germany}
\email{\authormark{*}michael.riesch@tum.de} 



\begin{abstract}
  We present a novel method to solve the Maxwell-Liouville-von Neumann (MLN) equations in an accurate and efficient way without invoking the rotating wave approximation (RWA). The method is a combination of two established concepts, namely the operator splitting method as well as the adjoint representation of the Lie algebra SU($N$) (or pseudospin representation). The former concept ensures the accuracy of the approach, but is computationally expensive. The latter concept provides an efficient representation of the problem and two optimization possibilities. We have implemented and verified both optimization approaches and demonstrate that substantial speedup can be achieved.
\end{abstract}





\section{Introduction}

The Maxwell-Liouville-von Neumann (MLN) equations describe the interaction of electromagnetic fields with quantum mechanical systems and are an important tool in nonlinear optics, for example to model the dynamics of quantum cascade lasers (QCLs)~\cite{jirauschek2014} or quantum well structures~\cite{dietze2011}. Usually, numerical methods are required to solve the MLN equations. The rotating wave approximation (RWA) is commonly used in order to save computation time but is avoided in this work since it omits certain features of the solution.

Several numerical methods to solve the MLN equations without invoking the RWA have been published in the last decades, starting with the pioneering work by Ziolkowski et al.~\cite{ziolkowski1995}. This work uses the finite-difference time-domain (FDTD) method for Maxwell's equations as well as the Crank-Nicholson scheme and a predictor-corrector (PC) method for the Liouville-von Neumann equations. In the latter, only two energy levels are considered (in this case, the equations are commonly called Maxwell-Bloch equations).

In the work by Slavcheva et al.~\cite{slavcheva2002, slavcheva2003}, this approach was extended to cope with more than two energy levels. The methods in \cite{ziolkowski1995, slavcheva2002, slavcheva2003} use the adjoint representation of the Lie algebra SU($N$) for $N$ energy levels (also called pseudospin representation) and hence eliminate redundant computations and memory requirements. Quantum mechanical operators such as the density matrix are Hermitian, therefore it is sufficient to calculate and store only the half of the off-diagonal elements. Additionally, the trace of the density matrix must equal 1, so the diagonal elements can be expressed using only $N-1$ real quantities. Regarding the numerical method, the predictor-corrector method is also computationally efficient.

However, Bid\'egaray et al.~\cite{bidegaray2001} proved that this method may produce unrealistic results when applied to multi-level systems. As an alternative that provides long-term stability in the simulations, an operator splitting approach was suggested~\cite{bidegaray2003, saut2006}. The major drawback of the operator splitting technique is the required computational effort to calculate the matrix exponentials at every time step. Different ways to perform this calculation efficiently can be found in literature: The approximation used in~\cite{bidegaray2001, bidegaray2003, saut2006}, employing the Expokit\cite{sidje1998} software package and replacing the FDTD with the pseudo-spectral time-domain (PSTD) method~\cite{marskar2011}, the Magnus expansion via Sylvester's formula~\cite{hailu2016}, the scaling and squaring method as well as a Krylov subspace method~\cite{guduff2017}, and diagonalization of the matrix~\cite{weninger2013}.

Although several research groups used the adjoint representation~\cite{ziolkowski1995, slavcheva2002, slavcheva2003, hailu2016, mathisen2016} or the operator splitting technique~\cite{bidegaray2001, bidegaray2003, saut2006, marskar2011, weninger2013}, there is -- to the best of our knowledge -- no approach that combines both concepts. In the work at hand we present such a combination and describe two optimization possibilities for the matrix exponential calculations.

In the following, we describe the Maxwell-Liouville-von Neumann equations (Section~\ref{sec:mln}) and the Liouville-von Neumann equation in the adjoint representation (Section~\ref{sec:adjoint}). In Section~\ref{sec:numerical} we focus on the numerical treatment of the Liouville-von Neumann equation. The resulting method can be used in combination with both FDTD and PSTD, therefore the treatment of Maxwell's equations shall receive less attention in the scope of this work. The verification of the presented methods and their computational performance are discussed in Section~\ref{sec:verification} using two test cases. Finally, we summarize the performance improvements of the presented methods and give an outlook on future work.

\section{The Maxwell-Liouville-von Neumann Equations}
\label{sec:mln}

In the work at hand, we consider one-dimensional Maxwell's equations for the electric and magnetic field components $E_z\left(x, t\right)$ and $H_y\left(x, t\right)$,
\begin{subequations}
\begin{align}
\partial_t E_z &= \epsilon^{-1} \left( -\sigma E_z - \partial_t P_z + \partial_x H_y \right),\\
\partial_t H_y &= \mu^{-1} \partial_x E_z,
\end{align}
\label{eq:maxwell}
\end{subequations}
where $x$ is the propagation direction, $y$ and $z$ denote the transversal coordinates, and $t$ is time. In Maxwell's equations the conductivity $\sigma$, permittivity $\epsilon$, and permeability $\mu$ of the active region material as well as the polarization term $P_z\left(x, t\right)$ are taken into account.

The second part of the MLN equations is the Liouville-von Neumann equation. It is used to determine the behavior of the quantum mechanical multilevel systems along the propagation direction (described by the density operator $\hat{\rho}\left(x, t\right)$, which can be written as $N \times N$ matrix for $N$ discrete energy levels). The equation reads
\begin{equation}
\partial_t \hat{\rho} = \mathcal{L} \left( \hat{\rho} \right) + \mathcal{G} \left( \hat{\rho} \right) = -\mathrm{i}\hslash^{-1} \left[ \hat{H}, \hat{\rho} \right] + \mathcal{G} \left( \hat{\rho} \right),
\label{eq:liouville}
\end{equation}
where the right hand side consists of the relaxation superoperator $\mathcal{G} \left( \hat{\rho} \right)$ (that may include e.g.\ scattering processes) and the Liouvillian $\mathcal{L} \left( \hat{\rho} \right)$. Here, $\hslash$ denotes the reduced Planck constant. The Hamiltonian $\hat{H}\left( x, t \right) = \hat{H}_0 + \hat{H}_{\mathrm{I}}\left( x, t \right)$ consists of a time-independent part $\hat{H}_0$ and a time-dependent interaction part $\hat{H}_{\mathrm{I}}\left( x, t \right) = -\hat{\mu} E_z\left( x, t \right)$, where $\hat{\mu}$ is the dipole moment operator. It is practical to define a corresponding pair of Liouvillian superoperators $\mathcal{L}_0 \left( \hat{\rho} \right)$ and $\mathcal{L}_{\mathrm{I}} \left( \hat{\rho} \right)$. We assume that only the electric field component $E_z\left( x, t \right)$ is relevant for the interaction process.

Finally, the polarization term $P_z\left( x, t \right)$ in Eq.~(\ref{eq:maxwell}) (or its derivative, respectively) is calculated as
\begin{equation}
\partial_t P_z = N_a \trace \left\{ \hat{\mu} \partial_t \hat{\rho} \right\} = N_a \trace \left\{ \hat{\mu} \mathcal{L} \left( \hat{\rho} \right) + \hat{\mu} \mathcal{G} \left( \hat{\rho} \right) \right\},
\label{eq:polarization1}
\end{equation}
where $N_a$ is the density of quantum mechanical particles in the system. Since
\begin{equation}
\trace \left\{ \hat{\mu} \left[ \hat{\mu}, \hat{\rho} \right] \right\} = \trace \left\{ \hat{\mu} \hat{\mu} \hat{\rho} \right\} - \trace \left\{ \hat{\mu} \hat{\rho} \hat{\mu} \right\} = \trace \left\{ \hat{\mu} \hat{\rho} \hat{\mu} \right\} - \trace \left\{ \hat{\mu} \hat{\rho} \hat{\mu} \right\} = 0,
\end{equation}
one can write Eq.~(\ref{eq:polarization1}) as
\begin{equation}
\partial_t P_z = N_a \trace \left\{ \hat{\mu} \mathcal{L}_0 \left( \hat{\rho} \right) + \hat{\mu} \mathcal{G} \left( \hat{\rho} \right) \right\} = N_a \trace \left\{ \mathcal{L}_0 \left( \hat{\rho} \right) \hat{\mu} + \mathcal{G} \left( \hat{\rho} \right) \hat{\mu} \right\},
\label{eq:polarization2}
\end{equation}
where $\mathcal{L}_0 \left( \hat{\rho} \right)$ represents the time-independent part of the Liouvillian.

\section{The Liouville-von Neumann Equation in Adjoint Representation}
\label{sec:adjoint}

The adjoint representation (also called pseudospin representation) is one of the most efficient ways to describe the density matrix, since only the non-redundant $N^2 - 1$ real elements are considered~\cite{hioe1981}. It is practical to write them as a vector $\vec{d}$. The density matrix $\hat{\rho}$ can be composed as
\begin{equation}
\hat{\rho} = N^{-1} \hat{I} + \frac{1}{2} \sum_{j=1}^{N^2 - 1} d_j \hat s_j,
\label{eq:comp_rho}
\end{equation}
where $\hat{I}$ is the $N \times N$ identity matrix, $\hat s_j$ are generators of the Lie algebra of SU($N$), and the vector elements are defined as $d_j \coloneqq \trace \left\{ \hat \rho \hat s_j \right\}$. The generators are traceless Hermitian $N \times N$ matrices and fulfill the condition $\trace \left\{ \hat s_j \hat s_k \right\} = 2 \delta_{jk}$, where $\delta_{jk}$ denotes the Kronecker delta. One possible choice~\cite{hioe1981} for the generators $\hat s = \left\{ \hat u_{12}, \dots, \hat v_{12}, \dots, \hat w_1, \dots, \hat w_{N - 1} \right\}$ consists of $N \left( N - 1 \right)/2$ generator pairs $\hat u_{jk} \coloneqq \hat t_{jk} + \hat t_{kj}$, $\hat v_{jk} \coloneqq -\mathrm{i} \left( \hat t_{jk} - \hat t_{kj} \right)$,
and $N - 1$ generators
\begin{equation}
  \hat w_{l} \coloneqq -\sqrt{\frac{2}{l \left( l + 1 \right)}} \left( \hat t_{11} + \hat t_{22} + \dots + \hat t_{ll} - l \hat t_{l+1, l+1} \right),
\end{equation}
where the indices satisfy $1 \leq j < k \leq N$ and $1 \leq l \leq N - 1$ and the transition-projection operators are defined as $\hat t_{jk} \coloneqq \ket{j}\bra{k}$. For $N=2$ and $N=3$ these generators produce the Pauli matrices and the Gell-Mann matrices, respectively.

In order to transform the Liouville-von Neumann equation to the adjoint representation, Eq. (\ref{eq:comp_rho}) is inserted into Eq. (\ref{eq:liouville}). Subsequently, the Frobenius inner product\footnote{The Frobenius inner product is defined as $\left< \hat a, \hat b \right>_F = \trace \left\{ \hat a \hat b^\dagger \right\}$. Note that $\left< \cdot, \hat s_k \right>_F = \trace \left\{ \cdot \hat s_k \right\}$, since the generators are Hermitian.} $\left< \cdot, \hat s_k \right>_F$ is applied to the result. The left hand side of Eq.~(\ref{eq:liouville}) then reads
\begin{equation}
\trace \left\{ \partial_t \hat{\rho} \hat s_k \right\} = \trace \left\{ \frac{1}{2} \sum_{j=1}^{N^2 - 1} \partial_t d_j \hat s_j \hat s_k \right\} = \frac{1}{2} \sum_{j=1}^{N^2 - 1} \partial_t d_j \trace \left\{ \hat s_j \hat s_k \right\} = \partial_t d_k.
\end{equation}
Since both superoperators are linear, we can write
\begin{subequations}
\begin{align}
  & \trace \left\{ \mathcal{L} \left( \hat \rho \right) \hat s_k + \mathcal{G} \left( \hat \rho \right) \hat s_k \right\} = \trace \left\{ \mathcal{L} \left( \hat \rho \right) \hat s_k \right\} + \trace \left\{ \mathcal{G} \left( \hat \rho \right) \hat s_k \right\},\\
  & \trace \left\{ \mathcal{L} \left( \hat \rho \right) \hat s_k \right\} = \underbrace{ \trace \left\{ N^{-1} \mathcal{L} \left( \hat{I} \right) \hat s_k \right\}}_{= 0} + \sum_{j=1}^{N^2 - 1} \frac{1}{2} \trace \left\{ \mathcal{L} \left( \hat s_j \right) \hat s_k \right\} d_j,\\
  & \trace \left\{ \mathcal{G} \left( \hat \rho \right) \hat s_k \right\} = \trace \left\{ N^{-1} \mathcal{G} \left( \hat{I} \right) \hat s_k \right\} + \sum_{j=1}^{N^2 - 1} \frac{1}{2} \trace \left\{ \mathcal{G} \left( \hat s_j \right) \hat s_k \right\} d_j
\end{align}
\end{subequations}
for the right hand side. As a result, the Liouville-von Neumann equation can be transformed to
\begin{equation}
\partial_t \vec{d} = \left( \mat{L} + \mat{G} \right) \vec{d} + \vec{d}^{\mathrm{eq}},
\label{eq:liouville-adj1}
\end{equation}
where the elements of the $N \times N$ matrices $\mat{L}$ and $\mat{G}$ are
\begin{subequations}
\begin{align}
l_{jk} &\coloneqq \frac{1}{2} \trace \left\{ \mathcal{L} \left( \hat s_k \right) \hat s_j \right\} = -\frac{\mathrm{i}}{2\hslash} \trace \left\{ \left[ \hat H, \hat s_k \right] \hat s_j \right\} = \frac{\mathrm{i}}{2\hslash} \trace \left\{ \hat H\left[\hat s_j, \hat s_k\right] \right\},\\
g_{jk} &\coloneqq \frac{1}{2} \trace \left\{ \mathcal{G} \left( \hat s_k \right) \hat s_j \right\},
\end{align}
\label{eq:elems1}
\end{subequations}
respectively, and the equilibrium vector $\vec{d}^{\mathrm{eq}}$ has the elements
\begin{equation}
d_{j}^{\mathrm{eq}} \coloneqq N^{-1} \trace \left\{ \mathcal{G} \left( \hat I \right) \hat s_j \right\}.
\end{equation}
With the separation of the Liouvillian in mind, we can split the matrix $\left( \mat{L} + \mat{G} \right)$ into a time-independent matrix $\mat{M}$ and a time-dependent part $\mat{U} E_z$, whose elements are
\begin{subequations}
\begin{align}
m_{jk} &\coloneqq \frac{\mathrm{i}}{2\hslash} \trace \left\{ \hat H_0\left[\hat s_j, \hat s_k\right] \right\} + g_{jk},\\
u_{jk} &\coloneqq \frac{\mathrm{i}}{2\hslash} \trace \left\{ -\hat \mu \left[\hat s_j, \hat s_k\right] \right\} = -\frac{\mathrm{i}}{2\hslash} \trace \left\{ \hat \mu \left[\hat s_j, \hat s_k\right] \right\},
\end{align}
\label{eq:elems2}
\end{subequations}
respectively, and write Eq.~(\ref{eq:liouville-adj1}) as
\begin{equation}
\partial_t \vec{d} = \left(\mat{M} + \mat{U} E_z \right) \vec{d} + \vec{d}^{\mathrm{eq}},
\label{eq:liouville-adj2}
\end{equation}
which is the Liouville-von Neumann equation in the adjoint representation.

Analogously, the derivative of the polarization must be expressed as function of the vector $\vec{d}$. In order to do so, we transform the dipole moment operator to a vector $\vec{v}$
\begin{equation}
  \hat \mu = \frac{1}{2} \sum_{k = 1}^{N^2 - 1} v_k \hat s_k \coloneqq \frac{1}{2} \sum_{k = 1}^{N^2 - 1} \trace \left\{ \hat \mu \hat s_k \right\} \hat s_k
\label{eq:comp_mu}
\end{equation}
using the generators as basis\footnote{This is possible since the generators span all traceless Hermitian $N \times N$ matrices. Due to the invariance of the Liouvillian superoperator, the main diagonal of every Hamiltonian (and consequently of the dipole moment operator) can be shifted so that the matrix becomes traceless.} and insert Eqs.~(\ref{eq:comp_rho}) and (\ref{eq:comp_mu}) into Eq.~(\ref{eq:polarization2}). The argument of the trace function in Eq.~(\ref{eq:polarization2}) then reads
\begin{equation}
\begin{split}
\left[ \mathcal{L}_0 \left( \hat{\rho} \right) + \mathcal{G} \left( \hat{\rho} \right) \right] \hat{\mu} &= \left[ \frac{1}{2} \sum_{j=1}^{N^2 - 1} d_j \mathcal{L}_0 \left(\hat s_j \right) + N^{-1} \mathcal{G} \left( \hat{I} \right) + \frac{1}{2} \sum_{j=1}^{N^2 - 1} d_j \mathcal{G} \left( \hat s_j \right) \right] \frac{1}{2} \sum_{k = 1}^{N^2 - 1} v_k \hat s_k \\
&= \frac{1}{2} \sum_{k = 1}^{N^2 - 1} v_k \left[ \sum_{j=1}^{N^2 - 1} d_j \frac{1}{2} \mathcal{L}_0 \left(\hat s_j \right) \hat s_k + N^{-1} \mathcal{G} \left( \hat{I} \right) \hat s_k + \sum_{j=1}^{N^2 - 1} d_j \frac{1}{2} \mathcal{G} \left( \hat s_j \right) \hat s_k \right]
\end{split}
\end{equation}
and can be simplified using the definitions in Eqs.~(\ref{eq:elems1}) and (\ref{eq:elems2}) once the trace function is applied. Then, the derivative of the polarization is calculated as
\begin{equation}
  \partial_t P_z = N_a \trace \left\{ \left[ \mathcal{L}_0 \left( \hat{\rho} \right) + \mathcal{G} \left( \hat{\rho} \right) \right] \hat{\mu} \right)\} = N_a \frac{1}{2} \sum_{k = 1}^{N^2 - 1} v_k \left( \sum_{j=1}^{N^2 - 1} m_{kj} d_j + d^{\mathrm{eq}}_k \right) = \frac{1}{2} N_a \vec{v}^T \left( \mat{M} \vec{d} + \vec{d}^{\mathrm{eq}} \right).
\end{equation}

\section{Numerical Treatment}
\label{sec:numerical}

\begin{figure}
\centering
\includegraphics[width=0.7\textwidth]{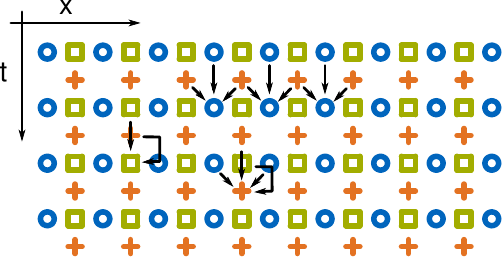}
\caption{Schematic of the discretization. The FDTD method uses a staggered grid for the discretization of the electric field $E_z$ (marked with crosses) and the magnetic field $H_y$ (circles). The density matrix $\hat \rho$ (or $\vec{d}$, respectively) and the resulting polarization $P_z$ is evaluated at the same location as the electric field, but shifted half a time step (squares).}
\label{fig:scheme}
\end{figure}

The achievements by Bid\'egaray et al.\ \cite{bidegaray2001, bidegaray2003} serve as basis for the numerical treatment of the equations described in the previous sections. The method in \cite{bidegaray2003} uses the finite-difference time-domain (FDTD) method \cite{taflove2005} to solve Maxwell's equations and an operator splitting approach for the Liouville-von Neumann equation. Figure~\ref{fig:scheme} provides a schematic overview. As already mentioned, we focus on the efficient numerical treatment of the Liouville-von Neumann equation in this study. Our efforts can be reused in an approach that uses a different method for Maxwell's equations, e.g.\ the pseudo-spectral time-domain (PSTD) method used in \cite{marskar2011}. Since the adjoint representation was not used in \cite{bidegaray2001, bidegaray2003}, we revisit the derivation of the operator splitting approach in the following. Subsequently, we describe two methods to compute the required matrix exponentials efficiently.

\subsection{Operator Splitting Approach}

The right hand side in Eq.~(\ref{eq:liouville-adj2}) can be split up in two operators $L_1$ (time-independent) and $L_2$ (time-dependent), yielding
\begin{equation}
\partial_t \vec{d} = L_1 \left( \vec{d} \right) + L_2 \left( \vec{d} \right).
\end{equation}
Subsequently, the differential equation is solved individually for each operator. For operator $L_1$, the solution is
\begin{equation}
\begin{split}
\partial_t \vec{d} = L_1 \left( \vec{d} \right) = \mat{M} \vec{d} + \vec{d}^{\mathrm{eq}} \rightarrow \vec{d}  &= \exp \left[ \mat{M} \left( t - t_0 \right) \right] \left[ \vec{d} \left( t_0 \right) + \mat{M}^{-1} \vec{d}^{\mathrm{eq}} \right] - \mat{M}^{-1} \vec{d}^{\mathrm{eq}}\\ &= \exp \left[ \mat{M} \left( t - t_0 \right) \right] \left[ \vec{d} \left( t_0 \right) + \vec{d}^{\mathrm{in}} \right] - \vec{d}^{\mathrm{in}},
\end{split}
\end{equation}
where $\vec{d}^{\mathrm{in}} \coloneqq \mat{M}^{-1} \vec{d}^{\mathrm{eq}}$ is the inhomogeneous part of the solution. The solution for operator $L_2$ can be determined as\footnote{Note that the Magnus expansion is not required since $\mat{U}$ is time-independent.}
\begin{equation}
\partial_t \vec{d} = L_2 \left( \vec{d} \right) = \mat{U} E_z \vec{d} \rightarrow \vec{d} = \exp \left[ \mat{U} \int_{t_0}^{t} E_z\left( \tau \right) \mathrm{d}\tau \right] \vec{d} \left( t_0 \right).
\end{equation}
The complete differential equation can be solved using the symmetric Strang splitting~\cite{strang1968}, i.e., by updating $\vec{d}$ first with the solution of operator $L_1$ for half a time step
\begin{equation}
\vec{d} \left(t_0 + \Delta t/2 \right) = \exp \left( \mat{M} \Delta t/2 \right) \left[ \vec{d} \left( t_0 \right) + \vec{d}^{\mathrm{in}} \right] - \vec{d}^{\mathrm{in}},
\end{equation}
then with the solution of operator $L_2$ for a full time step
\begin{equation}
\vec{d} \left(t_0 + \Delta t \right) = \exp \left[ \mat{U} \int_{t_0}^{t_0 + \Delta t} E_z\left( \tau \right) \mathrm{d}\tau \right] \vec{d} \left( t_0 \right) \approx \exp \left[ \mat{U} E_z\left( t_0 + \Delta t/2 \right) \Delta t \right] \vec{d} \left( t_0 \right),
\end{equation}
where the midpoint rule is applied as second order approximation for the integral, and finally with the solution of operator $L_1$ for half a time step again. The resulting operator splitting approach features second order accuracy as well and has the update rules
\begin{subequations}
\begin{align}
\vec{d}' &\leftarrow \exp \left( \mat{M} \Delta t/2 \right) \left( \vec{d}^{n - 1/2} + \vec{d}^{\mathrm{in}} \right) - \vec{d}^{\mathrm{in}},\\
\vec{d}'' &\leftarrow \exp \left( \mat{U} E_z^n \Delta t \right) \vec{d}',\\
\vec{d}^{n + 1/2} &\leftarrow \exp \left( \mat{M} \Delta t/2 \right) \left( \vec{d}'' + \vec{d}^{\mathrm{in}} \right) - \vec{d}^{\mathrm{in}},
\end{align}
\end{subequations}
where $\vec{d}^{n} = \vec{d} \left( n \Delta t \right)$ and $E_z^{n} = E_z \left( n \Delta t \right)$ represents the discretization at a given time step $n$.

In related work~\cite{bidegaray2003, saut2006, riesch2017b} the calculation of the matrix exponentials has been identified as the computational bottleneck. While the expression $\mat{A}_0 \coloneqq \exp \left( \mat{M} \Delta t/2 \right)$ is constant and can be calculated once (e.g.\ using the Pad\'e approximation), the interaction term $\mat{A}_{\mathrm{I}}^n \coloneqq \exp \left( \mat{U} E_z^n \Delta t \right)$ must be updated every time step. Therefore, the latter calculation should be optimized first.

\subsection{Efficient Computation of Matrix Exponentials}

By inspection of the elements $u_{jk}$ we can determine two properties of the matrix $\mat{U}$. The commutator $\hat C \coloneqq \left[ \hat s_j, \hat s_k \right]$ in Eq.~(\ref{eq:elems2}b) is antisymmetric, as a result the same holds for the elements $u_{jk}$. Furthermore, $\hat C$ is skew-Hermitian (which holds for every commutator of two Hermitian matrices). Then, the trace
\begin{equation}
  \trace \left\{ \hat \mu \hat C \right\} = \sum_{i=1}^{N} \mu_{ii} c_{ii} + \sum_{1 \leq j < k \leq N} \left(\mu_{jk}^* c_{jk} - \mu_{jk} c_{jk}^* \right)
\end{equation}
with $\mu_{jk} = \bra{j} \hat \mu \ket{k}$ and $c_{jk} = \bra{j} \hat C \ket{k}$ is purely imaginary, since $\mu_{ii}$ and $c_{ii}$ are real and purely imaginary, respectively, and $\mu_{jk}^* c_{jk} - \mu_{jk} c_{jk}^* = 2 \mathrm{i} \left[ \real \left( \mu_{jk} \right) \imag \left( c_{jk} \right) - \imag \left( \mu_{jk} \right) \real \left( c_{jk} \right) \right]$ is purely imaginary. As a result, the elements $u_{jk}$ in Eq.~(\ref{eq:elems2}b) are always real.

This fact leads to two optimization possibilities. First, a real antisymmetric matrix $\mat{U} = \mat{R}\mat{\Lambda}\mat{R}^\dagger$ can be diagonalized so that $\mat{\Lambda}$ is a diagonal matrix containing the (purely imaginary and pairwise complex conjugated) eigenvalues $\lambda_i$ and $\mat{R}$ is a unitary matrix consisting of the eigenvectors. The interaction term now reads
\begin{equation}
  \mat{A}_{\mathrm{I}}^n = \mat{R} \exp \left( \mat{\Lambda} E_z^n \Delta t \right) \mat{R}^\dagger = \mat{R}
  \begin{bmatrix}
    \exp \left( \lambda_1 E_z^n \Delta t \right) & \dots & 0 \\
      \vdots & \ddots & \vdots \\
      0 & \dots & \exp \left( \lambda_{N^2 - 1} E_z^n \Delta t \right) \\
  \end{bmatrix}
  \mat{R}^\dagger
\label{eq:exp2}
\end{equation}
and the cost of calculating the matrix exponential is reduced to two (complex) matrix multiplications and $N^2 - 1$ calls to the scalar exponential function. The matrix $\mat{R}$ as well as the eigenvalues remain constant and can be precalculated. This is one of the most accurate and efficient techniques for normal matrices~\cite{moler2003} (real antisymmetric matrices are always normal).

Second, this special case of matrix exponential allows analytic solutions. For two-level systems the exponential of $3 \times 3$ matrices has to be evaluated, which can be accomplished using Rodrigues' formula. This formula was generalized to arbitrary matrix dimensions~\cite{gallier2003}. As prerequisites, one has to determine the eigenvalues $\lambda_{2i-1,2i} = \pm \mathrm{i} \theta_i$ of $\mat{U}$ and set up a matrix $\mat{S}_i = \mat{T} \mat{\Theta}_i \mat{T}^{\mathrm{T}}$ for each pair of eigenvalues. The matrix
\begin{equation}
\mat{T} = \frac{1}{\sqrt{2}}
\begin{bmatrix}
\left( \vec{r}_1 + \vec{r}_2 \right) & -\mathrm{i} \left( \vec{r}_1 - \vec{r_2} \right) & \left( \vec{r}_3 + \vec{r}_4 \right) & -\mathrm{i} \left( \vec{r}_3 - \vec{r_4} \right) & \dots & \vec{r}_{N^2 - 1}
\end{bmatrix}
\end{equation}
is constructed using the eigenvectors $\vec{r}_{k}$ of $\mat{U}$ that correspond to the eigenvalues $\lambda_k$,\footnote{Note that $\mat{U}$ has an extra eigenvalue $\lambda_{N^2 - 1}=0$ besides the purely imaginary pairs in the odd-dimensional case.} the matrix $\mat{\Theta}_i$ is zero apart from the entries
\[
\begin{bmatrix}
  0 & -1\\
  1 & 0
\end{bmatrix}
\]
in the $(2i-1)$-th and $(2i)$-th row and column, respectively.\footnote{In the odd-dimensional case, the matrices $\mat{\Theta}_i$ are padded with zeros.} Then, the matrix exponential can be calculated as
\begin{equation}
\mat{A}_{\mathrm{I}}^n = \mat{I} + \sum_{i=1}^q \sin \left( \theta_i E_z^n \Delta t \right) \mat{S}_i + \left[ 1 - \cos \left( \theta_i E_z^n \Delta t \right) \right] \mat{S}_i^2,
\end{equation}
where $\mat{I}$ is the identity matrix and $q$ is the number of eigenvalue pairs.

\section{Verification and Performance Evaluation}
\label{sec:verification}

We implemented the diagonalization technique as well as the method using the generalized Rodrigues formula based on the mbsolve project~\cite{mbsolve-github}. This project has recently served as base for our operator splitting approach using the Eigen library~\cite{eigen-project} and the Pad\'e approximation for the matrix exponential calculations~\cite{riesch2017b}. The Pad\'e approximation approach serves as reference in terms of verification as well as performance in two test cases, which we describe in the following. Finally, we discuss the performance of each technique.

\subsection{Two-level system (Maxwell-Bloch case)}

\begin{figure}
\centering
\includegraphics[width=0.7\textwidth]{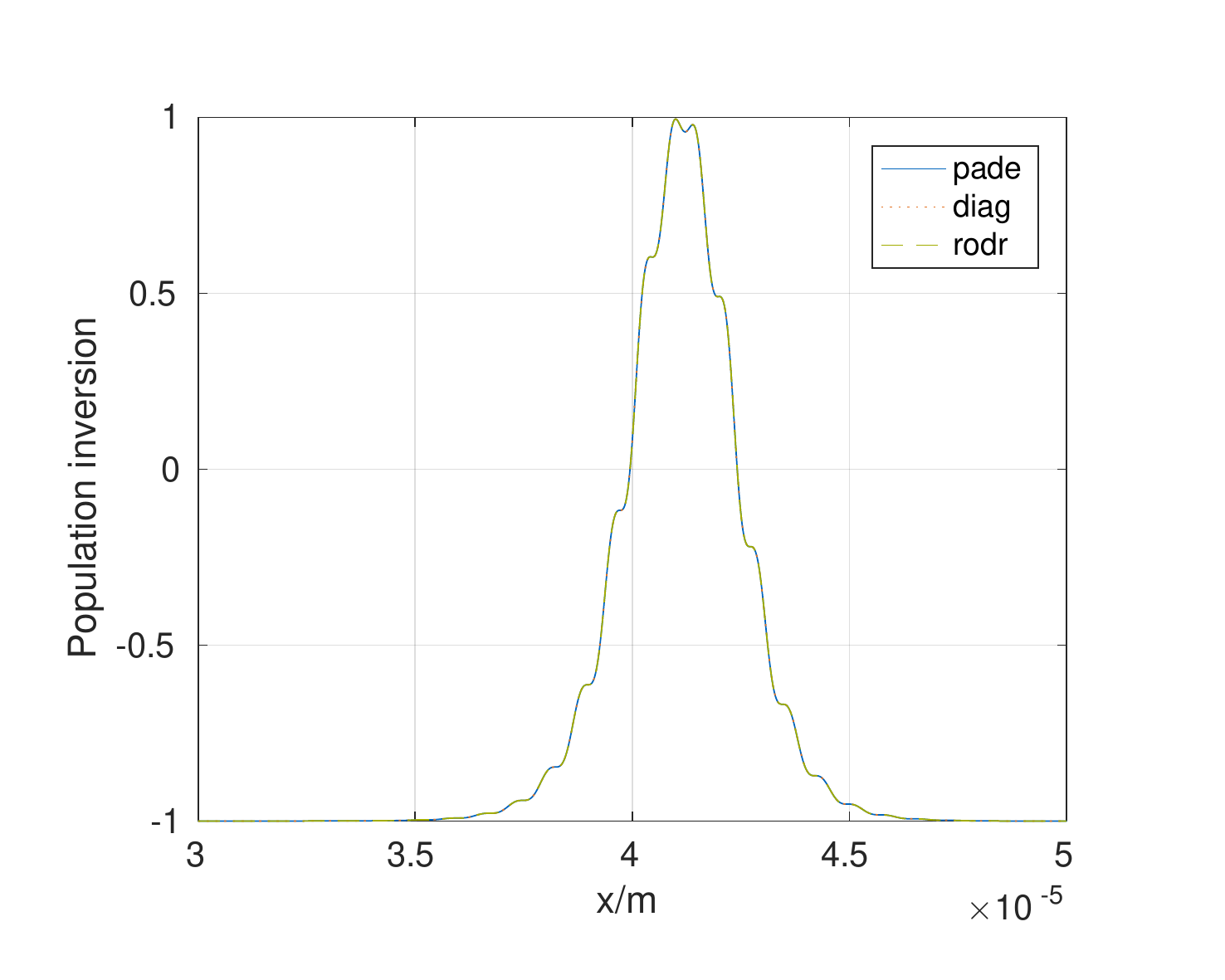}
\caption{Verification of the two-level simulation. The trace shows perfect agreement between the results of the Pad\'e approximation method (pade), the diagonalization approach (diag), and the generalized Rodrigues formula (rodr). The results of Ziolkowski et al.\ could be reproduced (cf.~\cite{ziolkowski1995}, Fig.~2).}
\label{fig:ziolkowski1995}
\end{figure}

The first test case is the self-induced transparency (SIT) simulation by Ziolkowski et al.~\cite{ziolkowski1995}. In this simulation, a two-level system is described with the Hamiltonian
\begin{equation}
  \hat{H} = \hat{H}_0 + \hat{H}_I = \hslash \omega_{12}
  \begin{bmatrix} -\frac{1}{2} & 0\\ 0 & \frac{1}{2} \end{bmatrix} -
  \begin{bmatrix} 0 & \mu_{12} \\ \mu_{12} & 0 \end{bmatrix} E_z,
\end{equation}
where $\omega_{12} = \SI{4\pi e14}{\per\second}$ is the transition frequency and $\mu_{12} = \SI{e-29}{\ampere\second\meter}$ is the dipole moment between the two levels, and the relaxation superoperator
\begin{equation}
  \mathcal{G} \left( \hat \rho \right) =
  \begin{bmatrix} T_1^{-1} \rho_{22} & -T_2^{-1} \rho_{12}\\ -T_2^{-1} \rho_{21} & -T_1^{-1} \rho_{22} \end{bmatrix},
\end{equation}
where $T_1 = \SI{e-10}{\per\second}$ and $T_2  = \SI{e-10}{\per\second}$ are decay terms (upper level lifetime and dephasing time, respectively).\footnote{Note that a different sign convention is used for $\mu_{12}$ in~\cite{ziolkowski1995}. However, the convention has no effect on the result if applied consistently.} These components are transformed to adjoint representation as described in Section~\ref{sec:adjoint}. Additionally, a initial density matrix
\begin{equation}
\hat \rho_0 = \begin{bmatrix} 1 & 0\\ 0 & 0 \end{bmatrix}
\end{equation}
can be transformed to an initial condition $\vec{d}_0$.

The simulation was set up with \num{32768} spatial grid points, which corresponds to a spatial discretization size of \SI{4.578}{\nano\meter}. By setting the Courant number $C=\num{0.5}$, we chose a time step size of \SI{7.635e-18}{\second}. Then, the simulation was executed and the populations of both levels were recorded. A snapshot after \SI{187.5}{\femto\second} is depicted in Fig.~\ref{fig:ziolkowski1995}. It shows perfect agreement between the results of all methods and the data depicted in \cite{ziolkowski1995}.

\subsection{Three-level system}

\begin{figure}
\centering
\includegraphics[width=0.7\textwidth]{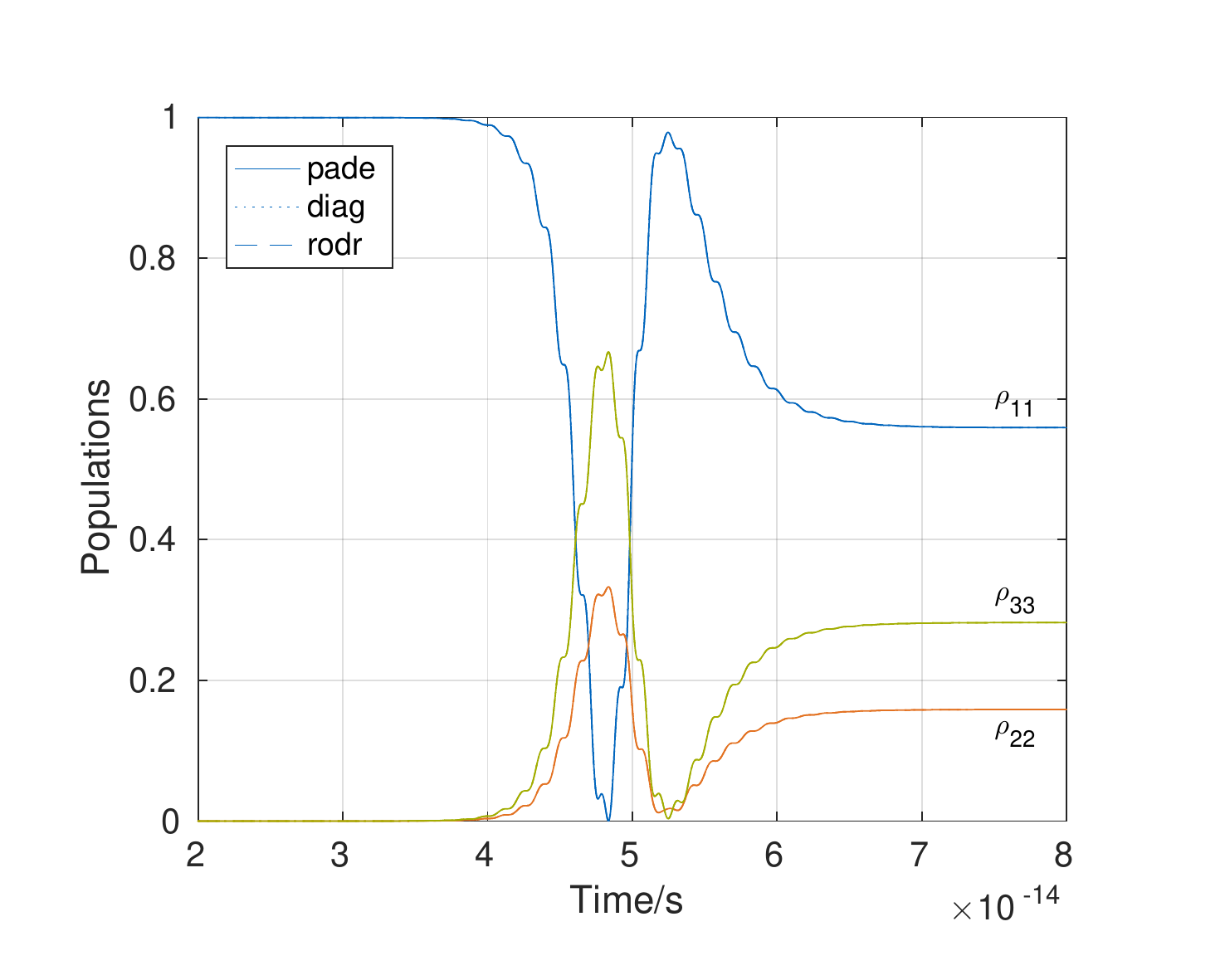}
\caption{Verification of the three-level simulation. Perfect agreement between the results of the Pad\'e  approximation method (pade), the diagonalization approach (diag), and the generalized Rodrigues formula (rodr) could be achieved. The results correspond to the findings by Song et al.~(cf.~\cite{song2005}, Fig.~3).}
\label{fig:song2005}
\end{figure}

The second test case incorporates the three-level setup presented by Song et al.~\cite{song2005}. The Hamiltonian is
\begin{equation}
  \hat{H} = \hat{H}_0 + \hat{H}_I = \hslash
  \begin{bmatrix} 0 & 0 & 0\\ 0 & \omega_2 & 0\\0 & 0 & \omega_3 \end{bmatrix}
  -
  \begin{bmatrix} 0 & \mu_{12} & \gamma \mu_{12}\\ \mu_{12} & 0 & 0\\
    \gamma \mu_{12} & 0 & 0
  \end{bmatrix} E_z,
\end{equation}
where $\omega_2 = \SI{2.372}{\per\second}$ and $\omega_3 = \SI{2.417}{\per\second}$ are the eigenfrequencies, $\mu_{12} = \SI{1.48e-29}{\ampere\second\meter}$ is the dipole moment between the levels 1 and 2, and $\gamma = \sqrt{2}$ is the ratio between the dipole moments. Furthermore, we define the relaxation superoperator as
\begin{equation}
  \mathcal{G} \left( \hat \rho \right) = T_1^{-1}
  \begin{bmatrix}
    \frac{1}{3} g_1 & - \rho_{12} & -\rho_{13}\\ -\rho_{21} & \frac{1}{3} g_1 - \left( \rho_{22} - \rho_{11} \right) & -\rho_{23}\\ -\rho_{31} & -\rho_{32} &  \frac{1}{3} g_1 - \left( \rho_{33} - \rho_{11} \right)\end{bmatrix},
\end{equation}
where $g_1 = \rho_{33} + \rho_{22} - 2 \rho_{11}$, and $T_1 = \SI{e-10}{\second}$ is the relaxation time. As initial condition, the density matrix
\begin{equation}
\hat \rho_0 = \begin{bmatrix} 1 & 0 & 0\\ 0 & 0 & 0\\ 0 & 0 & 0 \end{bmatrix}
\end{equation}
was transformed to the adjoint representation.

Using the choice of grid point size and time step size from the two-level test case, the results from \cite{song2005} could be reproduced. Figure~\ref{fig:song2005} shows the agreement of the different methods for this test case.

\subsection{Performance}

\begin{figure}
\centering
\includegraphics[width=0.7\textwidth]{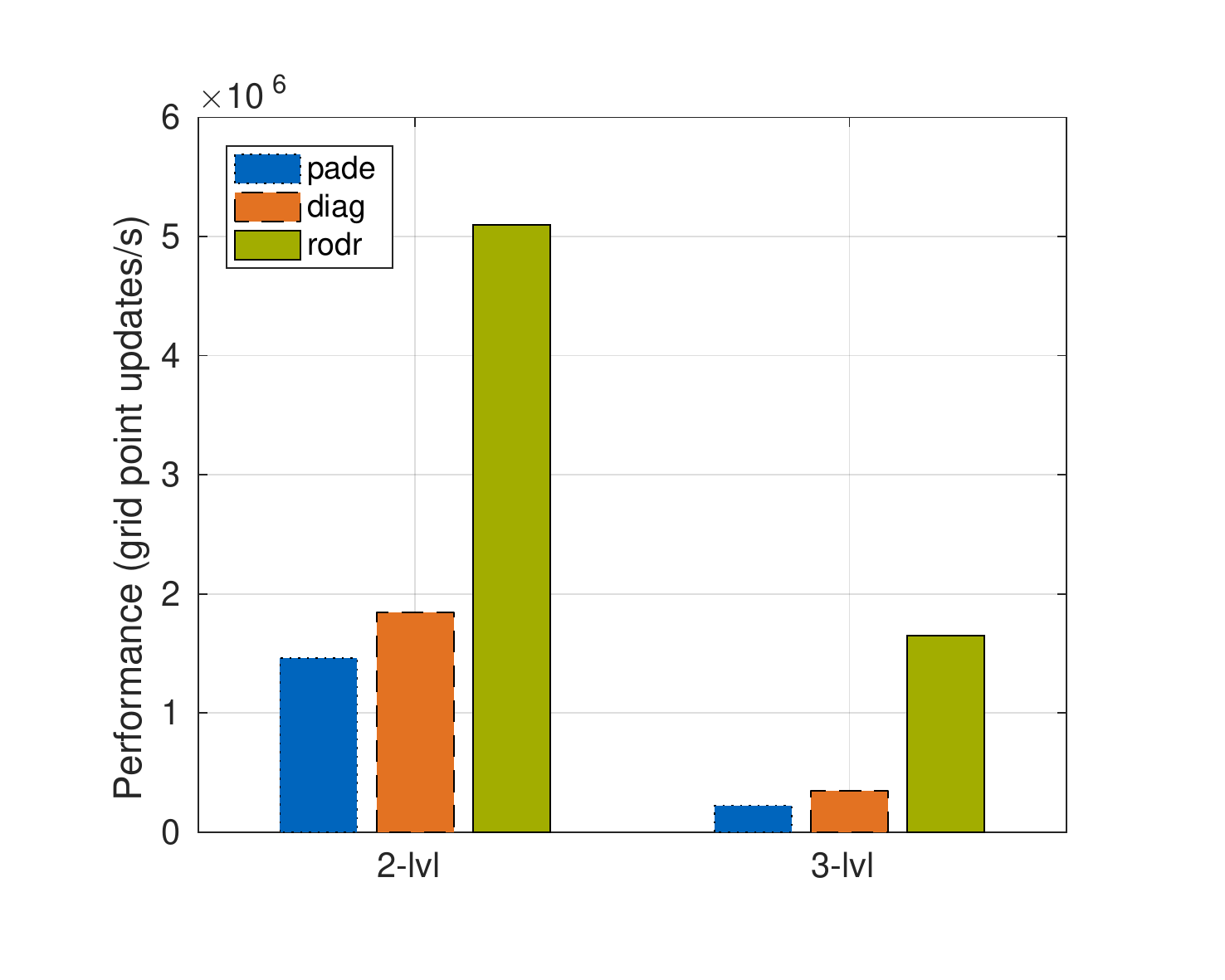}
\caption{Single-thread performance comparison of the Pad\'e approximation method (pade), the diagonalization approach (diag), and the generalized Rodrigues formula (rodr) applied to the two-level and the three-level test case.}
\label{fig:comparison}
\end{figure}

The simulations were executed on an Intel Xeon Processor E7-4870. Although the code is designed to run efficiently in parallel, only one thread was used since we wanted to measure the quality of a single-thread optimization. In order to ensure reproducibility, the measurements were repeated five times. The performance was determined as number of grid point updates per time unit, i.e.\ $P = N_xN_t/t_{\mathrm{exec}}$, where $N_x$ and $N_t$ are the number of spatial and temporal grid points, respectively, and $t_{\mathrm{exec}}$ is the measured execution time. This allows the metric to be used to compare the performance of problems with different sizes.

It should be noted that the execution time does not contain the time required to construct or delete data structures (which is negligible for the simulation setups in question). It does, however, contain the time required to store the result data, in particular the time to convert the vector $\vec{d}$ to the density matrix $\hat \rho$.

The performance values of the different methods applied to the two test cases are shown in Fig.~\ref{fig:comparison}. Compared to our baseline (the Pad\'e  approximation method), the diagonalization approach performs \num{1.26}x better for the two-level test case and \num{1.55}x better for the three-level test case. Using the generalized Rodrigues formula, we were able to achieve a speedup of \num{3.48}x and \num{7.35}x, respectively.

\section{Conclusion}

Our method solves the Maxwell-Liouville-von Neumann equations in an efficient and accurate way. It is a novel combination of two established concepts. The adjoint representation is most efficient in terms of eliminating the redundancy inherent to quantum mechanical operators and additionally leads to interesting properties of the resulting description. These properties are exploited in order to provide two efficient implementations of the operator splitting technique, which is accurate and stable, yet computationally expensive.

Both implementations are tested with the help of a two-level and a three-level test case and can reproduce the results found in related literature. Regarding the performance, the diagonalization approach yields a small improvement and the method using the generalized Rodrigues formula provides substantial speedup, at least at test cases with small level count.

We expect that both methods provide even larger performance improvements compared to standard methods (e.g.\ the Pad\'e approximation) in test cases with many energy levels. Also, the performance ratio between the diagonalization approach and the Rodrigues method may differ. However, the behavior of the methods applied to a many-level setup has to be investigated first.

Finally, the method is able to cope with a broad range of problems. However, there are two limitations, namely the restriction to one-dimensional models as well as the strict definition of the interaction Hamiltonian. In future work we will aim to generalize our method in order to overcome those limitations.

\section*{Supplementary Material}

See \cite{mbsolve-github} for source code, build instructions, and basic documentation. In this paper we used the development branch \texttt{riesch2017c}, the Eigen library version 3.3.4, and the Intel C++ compiler 17.0.

\section*{Funding}

German Research Foundation (DFG) (JI 115/4-1, JI 115/9-1).

\section*{Acknowledgments}

The authors gratefully acknowledge the Gauss Centre for Supercomputing e.V. (www.gauss-centre.eu) for funding this project by providing computing time on the GCS Supercomputer SuperMUC at Leibniz Supercomputing Centre (www.lrz.de). We thank Gabriela Slavcheva for the interesting discussion and her valuable input on the adjoint representation.

\end{document}